# VLBI Tracking of the JUICE Mission: Two Years of Cruise Phase Operations and Performance Analysis

O. J White [1], Guifre Molera Calves [1], Jasper Edwards [1], Giuseppe Cimo [2], Dominic Dirkx [3], on behalf of the PRIDE team.

[1] *School of Natural Sciences, University of Tasmania, Hobart, TAS, Australia*

[2] *Joint Institute for VLBI in Europe, Dwingeloo, The Netherlands*

[3] *Technical University of Delft, Delft, The Netherlands*

**Summary:** The JUpiter ICy moons Explorer (JUICE) mission, launched by the European Space Agency (ESA) in April 2023, represents one of the most ambitious deep space exploration endeavours targeting Jupiter's icy moons. This paper presents results from the Very Long Baseline Interferometry (VLBI) radio telescope tracking conducted by the University of Tasmania during the first two years of JUICE's cruise phase operations. We have conducted over 100 tracking sessions capturing the spacecraft across different orbital regimes as JUICE progresses through its complex cruise trajectory towards Jupiter. Our analysis focuses on three key areas: Doppler residual characterisation, mission performance indicator extraction, and solar wind scintillation pattern analysis (including space weather forecasting). UTAS measurements demonstrate the enhanced capabilities that VLBI networks provide for deep space mission support, particularly for precision orbit determination and spacecraft health diagnosis. The results showcase the UTAS VLBI array as a valuable complement to traditional tracking infrastructure, offering Southern Hemisphere coverage and enhanced geometric diversity for deep space missions.

## Introduction

The Very Long Baseline Interferometry (VLBI) observations conducted of the JUpiter ICy moons Explorer (JUICE) mission are part of the Planetary Radio Interferometry and Doppler Experiment (PRIDE). This technique has been developed over twenty years of VLBI and single-dish observations of various spacecraft, descended from the VLBI observations conducted of the European Space Agency's (ESA's) Huygens Titan probe [1]. It was an unprecedented global campaign utilising 17 radio telescopes to obtain Doppler and VLBI tracking data to enhance the Doppler Wind Experiment and to determine the probe's position in the celestial plane.

The technique has since been refined for experiments using ESA's Venus Express (VEX) and Mars Express (MEX) spacecraft. These have included radio occultation experiments of Venus' atmosphere, to reconstruct its vertical temperature and pressure profiles [2], and precise phase-referencing VLBI tracking of Mars Express during the 2013 of Phobos; VLBI imaging enabling determination of displacements from the a priori lateral position of MEX in right ascension and declination with milli-arcsecond precision [3]. Single-dish observations have also been utilised for the study of space weather through measurements of the interplanetary scintillation along the line of sight between spacecraft and the Earth [4].

The JUICE mission to the Jovian system aims to study Jupiter's icy moons Europa, Ganymede and Callisto in detail [5]. It is currently undergoing an eight-year cruise phase, after its launch in April 2023. A Lunar-Earth gravity assist was conducted in August 2024, and a flyby of Venus in August 2025. Future flybys of the Earth are planned for September 2026 and January 2029, before its arrival at the Jovian system in 2031.

Throughout the first two years of the cruise phase the University of Tasmania (UTAS) and the PRIDE team have conducted over 100 observations of JUICE. These included nine phase-referencing VLBI experiments in 2024 and 2025, from which JUICE was successfully imaged [6]. We have used the UTAS 30m Ceduna (Cd), 26m Hobart (Ho), and the 12m Hobart (Hb), Katherine (Ke), and Yarragadee (Yg) radio telescopes for these sessions.

This paper is divided into three sections; first, we detail the methodology used for this campaign, including the observation setup, data processing, and software pipeline. Next, we discuss the results, with focuses on the near-Earth commissioning phase, the Lunar-Earth gravity assist, the Venus flyby, and space weather analysis. Finally, we conclude with a discussion of further analysis and observations planned for the remainder of the cruise phase.

## Methodology

### Observation Setup

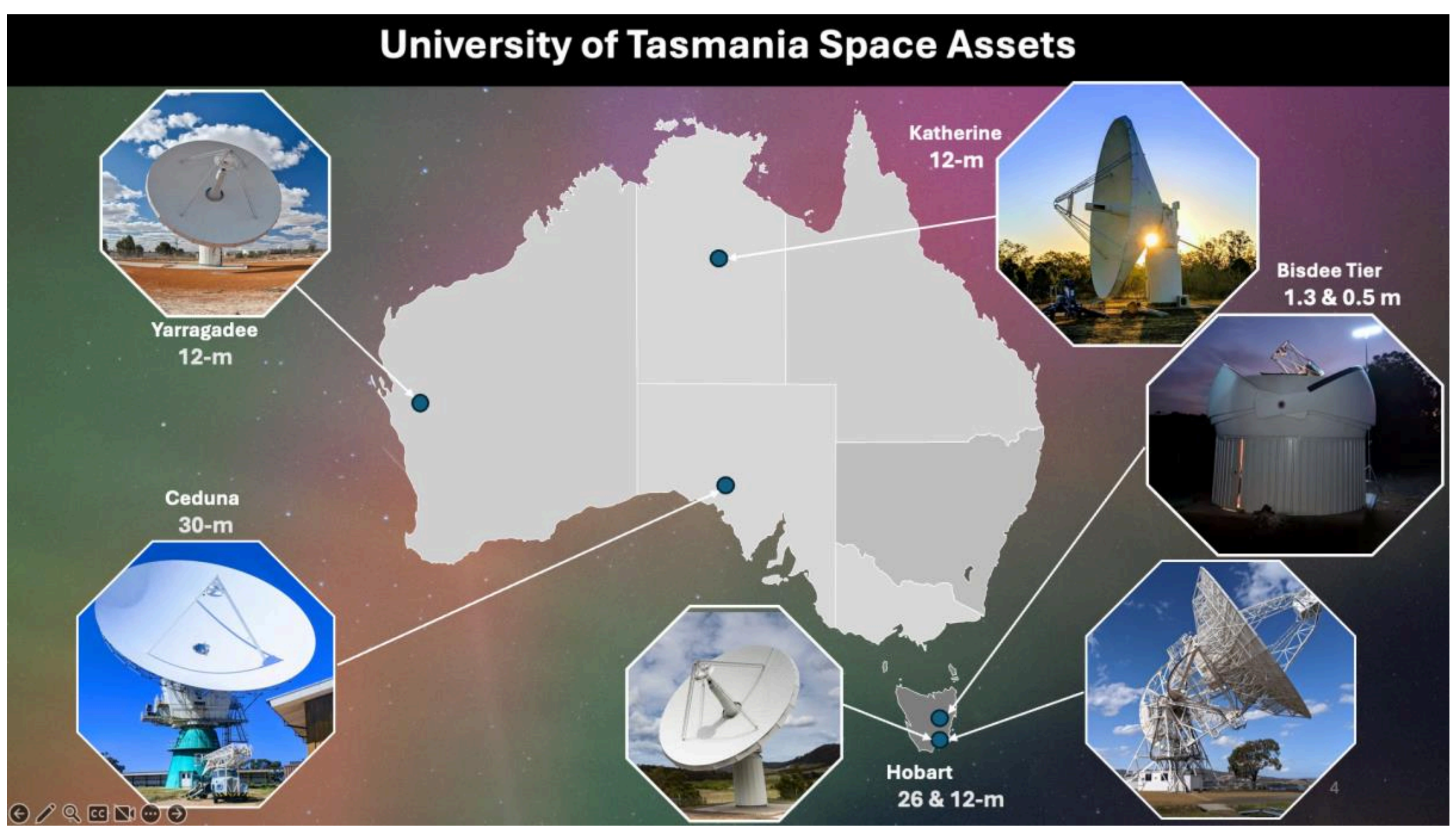

*Figure 1: Map of the UTAS optical and radio telescopes.*

Radio observatories worldwide are equipped with specialized hardware for interferometric observations, enabling the recording of wide-bandwidth signals from spacecraft and astronomical sources with exceptional timing precision and sensitivity. The University of Tasmania operates all sites with Digital BaseBand Converter 2 (DBBC2) or DBBC3 backends paired with high-capacity storage systems capable of recording data rates up to 32 Gbps. Recently, we have introduced more flexible Software-Defined Radio (SDR) systems that provide real-time feedback on signal acquisition and facilitate studies of space weather phenomena, including amplitude scintillation and frequency broadening; capabilities not supported by legacy systems. Both approaches, traditional VLBI hardware and the new Ettus Universal Software Radio Peripheral (USRP) devices, operate in parallel, ensuring redundancy and verification.

We have primarily conducted two types of observations, as detailed in [4]: single-station Doppler measurements and VLBI phase-referencing experiments. While the former involves single-dish observations, it can include multiple stations operating independently. In contrast, phase-referencing requires alternating observations between the spacecraft and a nearby calibrator radio source to enable astrometric measurements. Doppler measurements are primarily used for space weather studies and for validating predicted orbits against observed data; these results are subsequently processed with the *TUDAT* software [14] to refine orbital solutions for the PRIDE team. Phase-referencing, on the other hand, has been demonstrated to provide two-dimensional estimates of the spacecraft state vector in the plane of the sky. Doppler observations have been conducted regularly throughout the two-year mission, whereas phase-referencing has been reserved for key events, such as prior to the Lunar-Earth flyby and the Venus flyby. Figure 2 is a heliospheric map of the observations observed by the UTAS antennas. Table 1 details the VLBI observations.

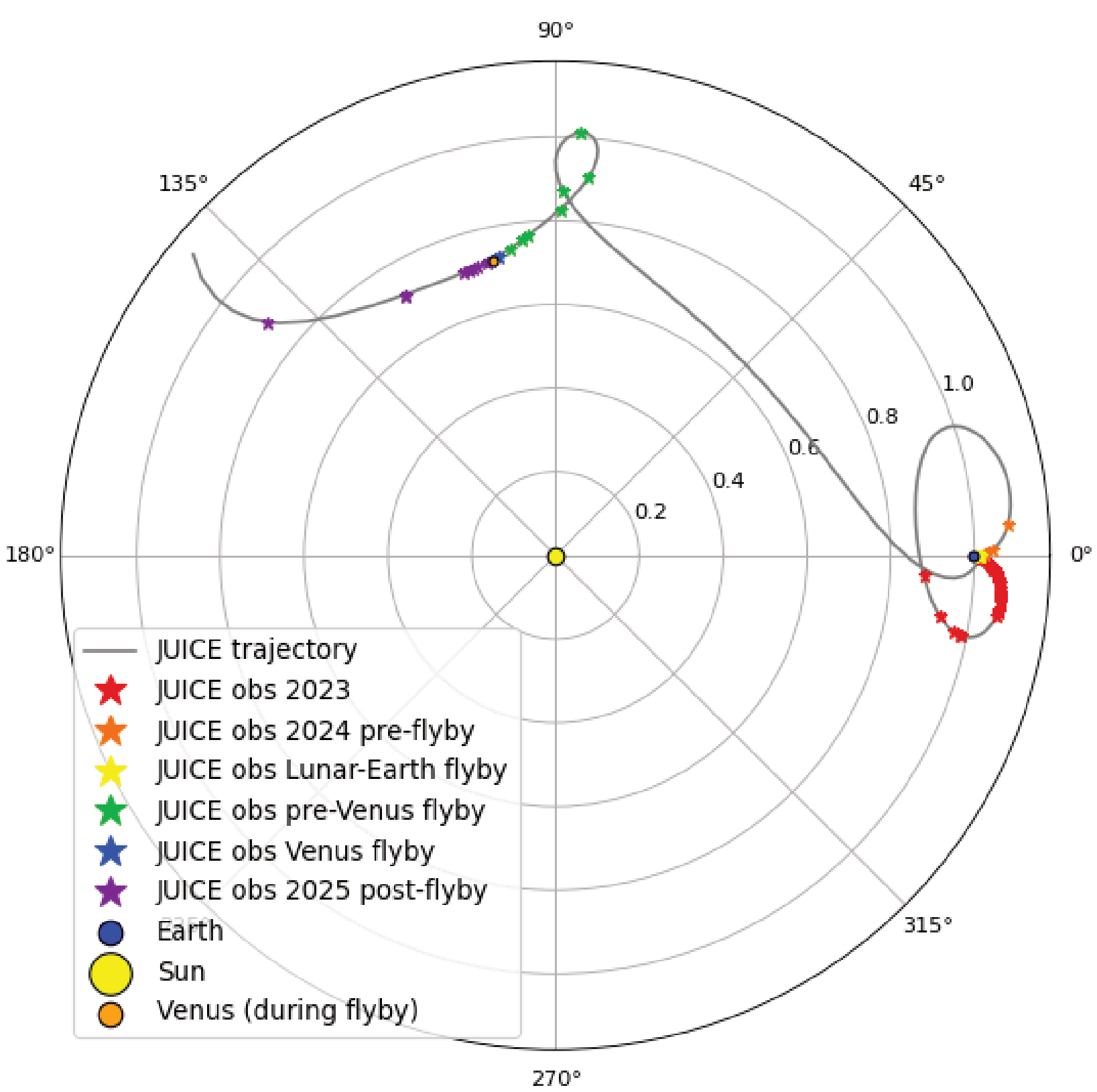

*Figure 2: Map of JUICE's trajectory relative to the Earth from the launch in April 2023 to December 2025. Dates of UTAS space weather and PRIDE VLBI observations are starred.*

*Table 1: Details of the nine PRIDE VLBI observations conducted at UTAS.*

| Epoch | Stations | Fringe Finder Sources | Phase Reference Sources |
|---|---|---|---|
| **02/07/2024** | Hb, Ke, Yg, Cd | J1924-2914 | J2211-1328, J2211-1150 |
| **03/08/2024** | | J1924-2914, J2258-2758 | J2229-0832, J2240-0836 |

| **13/08/2024** | Hb, Ke, Yg | J2258-2758 | J2229-0832 |
|---|---|---|---|
| **19/08/2024** | | J1924-2914 | J2211-1328 |
| **29/08/2025** | Hb, Ke, Yg, Cd | J0854+2006 | J0832+1832, J0839+1802 |
| **31/08/2025** | | J0854+2006 | J0832+1832, J0839+1802 |
| **05/09/2025** | Hb, Ke, Yg | J0854+2006 | J0908+1609, J0925+1658 |
| **06/09/2025** | | J0854+2006 | J0908+1609, J0925+1658 |
| **19/09/2025** | Hb, Ke, Yg, Cd | J1058+0133, J1041+0610, J1038+0512 | J1015+1227, J1025+1253 |

The tracking of the spacecraft has also been upgraded from previous PRIDE experiments. Original observations of PRIDE [4, 7] used the legacy tracking systems at the stations with repointing of the targets every 5 to 20 minutes depending on the spacecraft motions and the length of the scans recorded. Now the *SATRK* satellite tracking software is used to provide pointing commands to the telescopes in 10 second intervals for the single-station Doppler experiments, enabling more precise tracking and hence stronger SNRs. The VLBI observations continue to use predicted right ascension and declination positions updated in 20-minute intervals.

## Data Processing

The data processing pipeline for these experiments varied based on whether it was a Doppler tracking or a PRIDE VLBI phase-referencing session. For the phase-referencing sessions, we used the *Distributed FX* (*DiFX*) *software correlator* [8] to correlate the raw data, with *Difxcalc* [9] used to generate the delay models. For JUICE, this utilised the Duev near-field delay model [10], with the ephemerides generated from the JUICE SPICE kernel[1]. The correlated data was then compiled into a FITS-IDI file for analysis with the *Astronomical Image Processing System* (*AIPS*). This enabled phase calibration and imaging of the spacecraft, facilitating measurement of the offset of the spacecraft from phase centre to its apparent position.

The single-dish Doppler observations were processed using the *Spacecraft Doppler tracking* (*SDtracker*) *software* [11]. The first step was producing broadband spectra of the entire channel containing the spacecraft signal using *SWspectrometer*, typically with a 10 second integration time. Next, the spacecraft signal was tracked in frequency and used to generate phase polynomials in 20-minute scans. These phase polynomials were then used with *SCtracker*, which phase stopped the carrier tone and produced a binary file of the spectrum of width 2 kHz, centred on the expected tone. Finally, we applied a digital phase lock loop (PLL) to generate high precision frequency and phase detections at smaller timescales, typically 1 second integration time.

Scintillation analysis of the phase residuals is performed to determine the contributions of space weather activity. Each scan is visually inspected for phase jumps or imperfect unwrapping of the phase, and any such are disregarded. Suitable scans are then analysed further, first by evaluating the root mean square (RMS) of the phase fluctuations, then using a fast Fourier transform to produce a phase power spectrum. Both a windowed and an unwindowed spectra is used in the analysis; a windowed spectra is used to estimate the slope and relate it to a Kolmogorov spectrum. The unwindowed spectra is used to estimate the RMS of the scintillation and the system noise. Both the RMS of the scintillation and the slope of the Kolmogorov spectrum can be used to estimate properties of the solar wind [4].

[1] https://s2e2.cosmos.esa.int/bitbucket/projects/SPICE_KERNELS/repos/juice/browse/kernels/spk

# Results

## Near-Earth Commissioning Phase (NECP)

During the near-Earth commissioning phase of the mission, observations were conducted over more than 35 separate epochs, with one or more UTAS antennas recording with the DBBC or Ettus. Details of the phase scintillation analysis during this period are listed in Table 2.

This provided an opportunity for testing different observation configurations and modes of transmission. Figure 3 demonstrates a comparison of the two-way and one-way Doppler observation modes. The first is a scan from June 23$^{rd}$, 2023, operating in two-way mode, and the second is from July 6th, operating in one-way mode, both from Hb. While their SNRs are of the same order of magnitude, both the residual frequencies and phases from the one-way mode are on average a factor of 100 greater than two-way mode. This renders the phases unusable for analysis of the phase scintillation for space weather applications. Additionally, this reduces the precision of the frequency detections for use in orbital modelling applications.

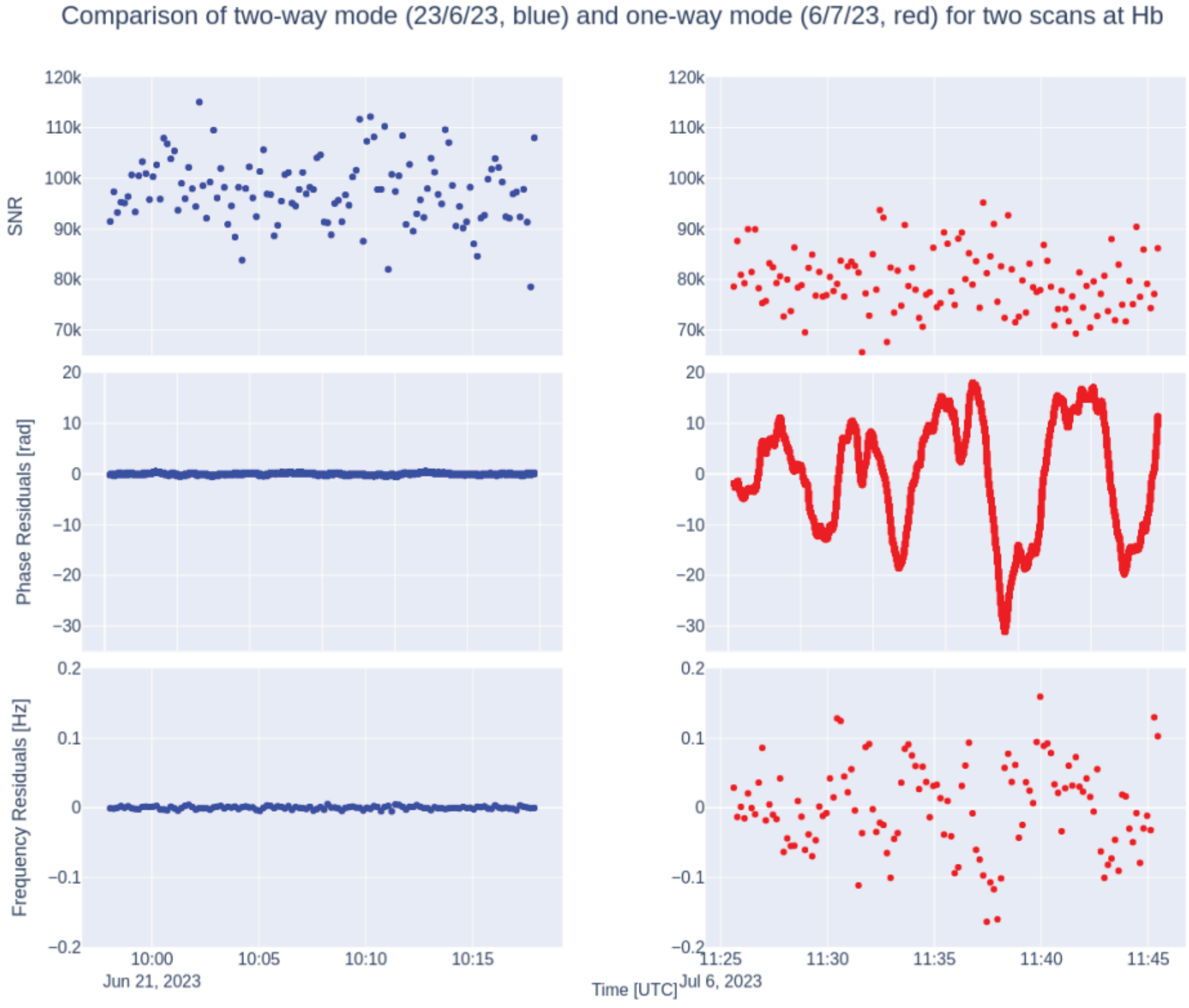

*Figure 3: Comparison of the two-way (left) and one-way (right) Doppler observation modes for two scans from separate observations. They have SNRs on the same order of magnitude, with the variation attributed to variance in the elevation of the antennas. The one-way mode shows a far inferior response in both the frequency and phase residuals*

.

Comparisons of observations using different polarisations were also conducted. JUICE transmits in left circular polarisation (LCP). The 12m antennas (Hb, Ke, Yg) each feature dual linear polarisation receivers, hence either can be used interchangeably for analysis. However, Cd and Ho feature circularly polarised receivers, so the left and right circular polarisation (RCP) properties of the signal can be analysed. Figure 4 shows a comparison of the SNR and residual frequencies for LCP and RCP at Ho on April 24th. The opposite polarisation (RCP) has a significantly reduced SNR as well as noisier frequency residuals.

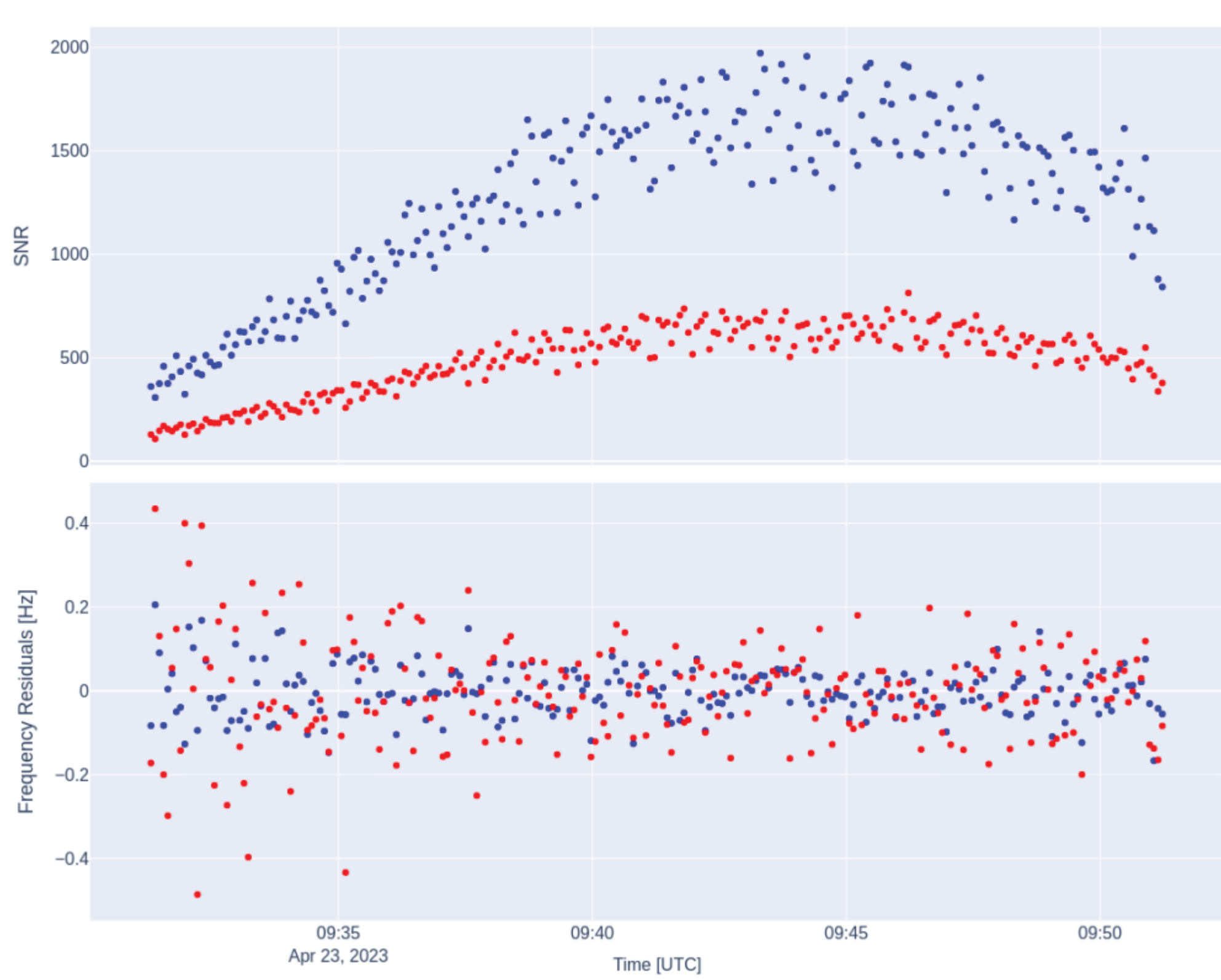

*Figure 4: Comparison of the LCP (blue) and RCP (red) responses for the same scan at Ho. JUICE typically transmits in LCP, hence the significantly reduced SNR in RCP, along with greater noise in the frequency residuals. In both cases, the frequency residuals decrease in time as the SNR increases.*

**Lunar-Earth Gravity Assist**

Prior to the Lunar-Earth gravity assist, multiple PRIDE VLBI phase-referencing observations were conducted at UTAS, on July 2$^{nd}$, August 3$^{rd}$, 13$^{th}$ and 19$^{th}$, 2024. As described in Table 1, the August 3$^{rd}$ observation used the Hb, Ke and Yg telescopes, with J1924-2914 and J2258-2758 used as fringe finder sources. J2229-0832 and J2240-0836 were the phase calibrator sources.

As discussed in the analysis of [6], the J2229-0832 scans were definitively detected. However, the J2240-0836 scans were too weak for successful correlation. The JUICE scans were also successfully correlated using the near-field Duev model. Images were successfully produced of the spacecraft and J2229-0832 using the Astronomical Image Processing System (AIPS) [1], shown in Figure 5 below. Further analysis of this dataset and additional experiments is ongoing.

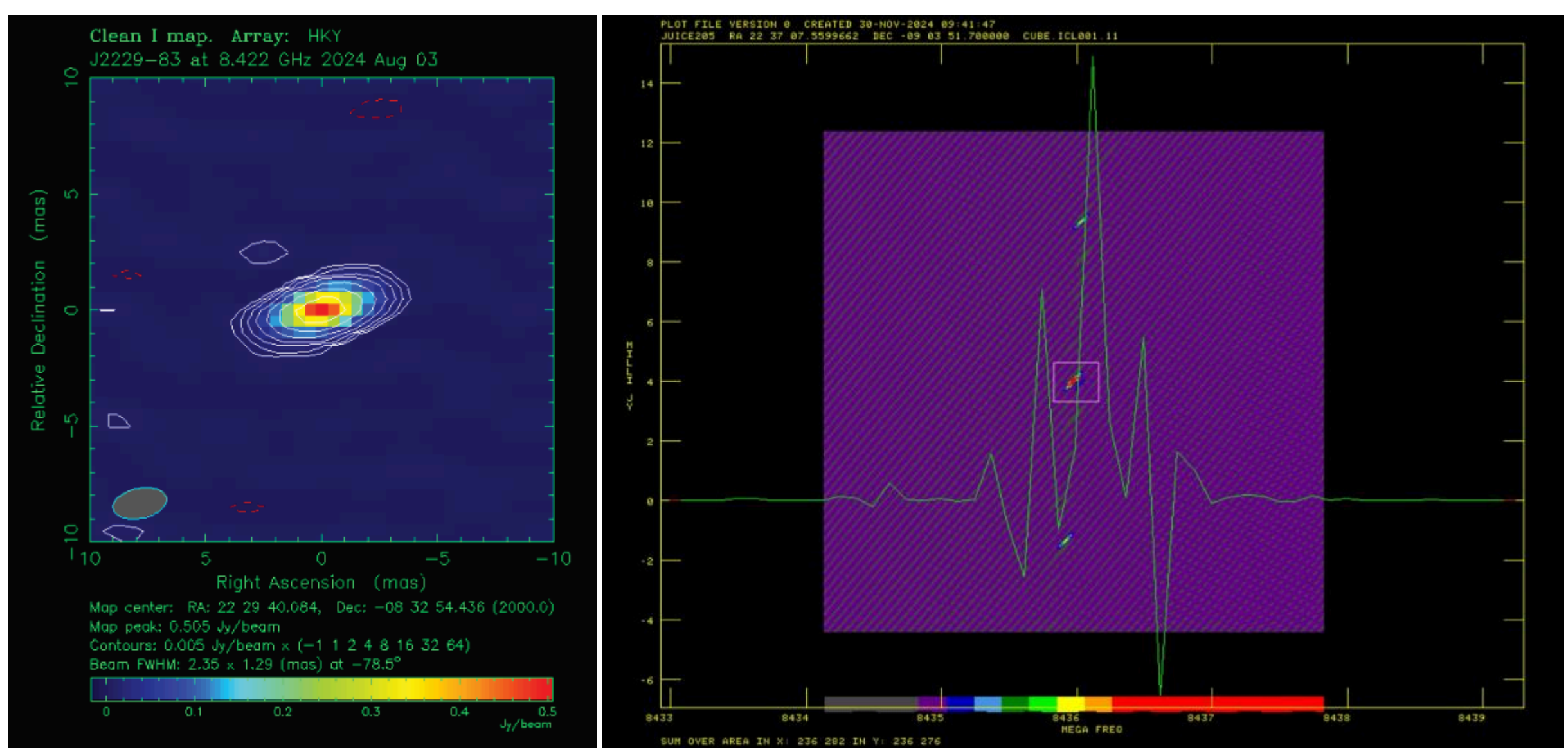

*Figure 5: Images of phase-reference source J2229-0832 (left) and JUICE (right) produced from the August 3rd, 2024, VLBI experiment [6]. JUICE is highlighted in the central box, with a frequency spectrum of the spacecraft signal overlaid.*

The Lunar-Earth gravity assist was composed of subsequent flybys of the Moon and the Earth. The Lunar flyby occurred on August 19th, 2024, and the Earth flyby on August 20th. Across this flyby period, JUICE had switched to its low-gain antenna due to the decreased power requirements. A PRIDE VLBI session was scheduled for the hours before the Lunar flyby; however, due to recording errors not enough data was collected for successful correlation and imaging. The Doppler tracking during the flyby was successful.

We tracked the Lunar flyby using the 26m Hartebeesthoek (Hh) antenna in South Africa, as this provided greater coverage than the UTAS antennas. As shown in Figure 6, we captured several events during the flyby. At 20:30 UTC the spacecraft switched from two-way communications to one-way in preparation of loss of contact due to Lunar occultation. Ingress occurred at 20:38, and the signal was reacquired during egress at 21:08. Two-way communications were restored at 21:13, and the closest approach to the Moon was captured at 21:16. Radio occultation analysis of the ingress period, 20:30 to 20:38, is ongoing.

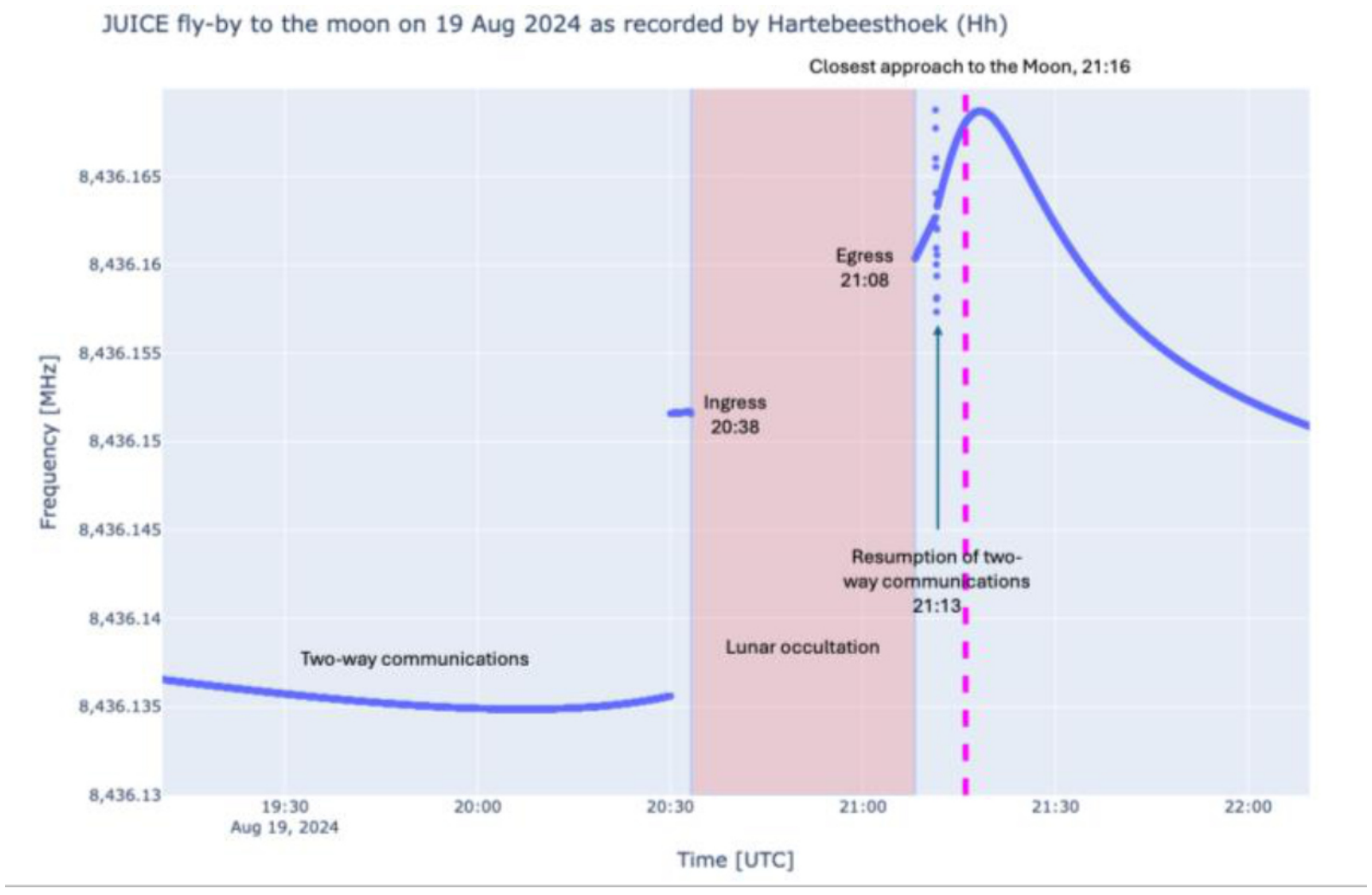

*Figure 6: Frequency detections of the JUICE flyby of the moon as observed by Hh, covering the Lunar occultation period and closest approach.*

Prior to the Earth flyby the higher elevation of JUICE over Australia enabled Hb, Ke and, Yg to also track JUICE, as shown in Figure 7 below. Unfortunately, due International Telecommunication Union regulations the spacecraft could not transmit at full power within 45,000 km of the Earth. Hence communications were cutoff at 19:28 on August 20th and did not recommence until 04:16 on the 21st. Thus, the spacecraft was unable to be tracked during the closest approach to the Earth at 21:57 on the 20th.

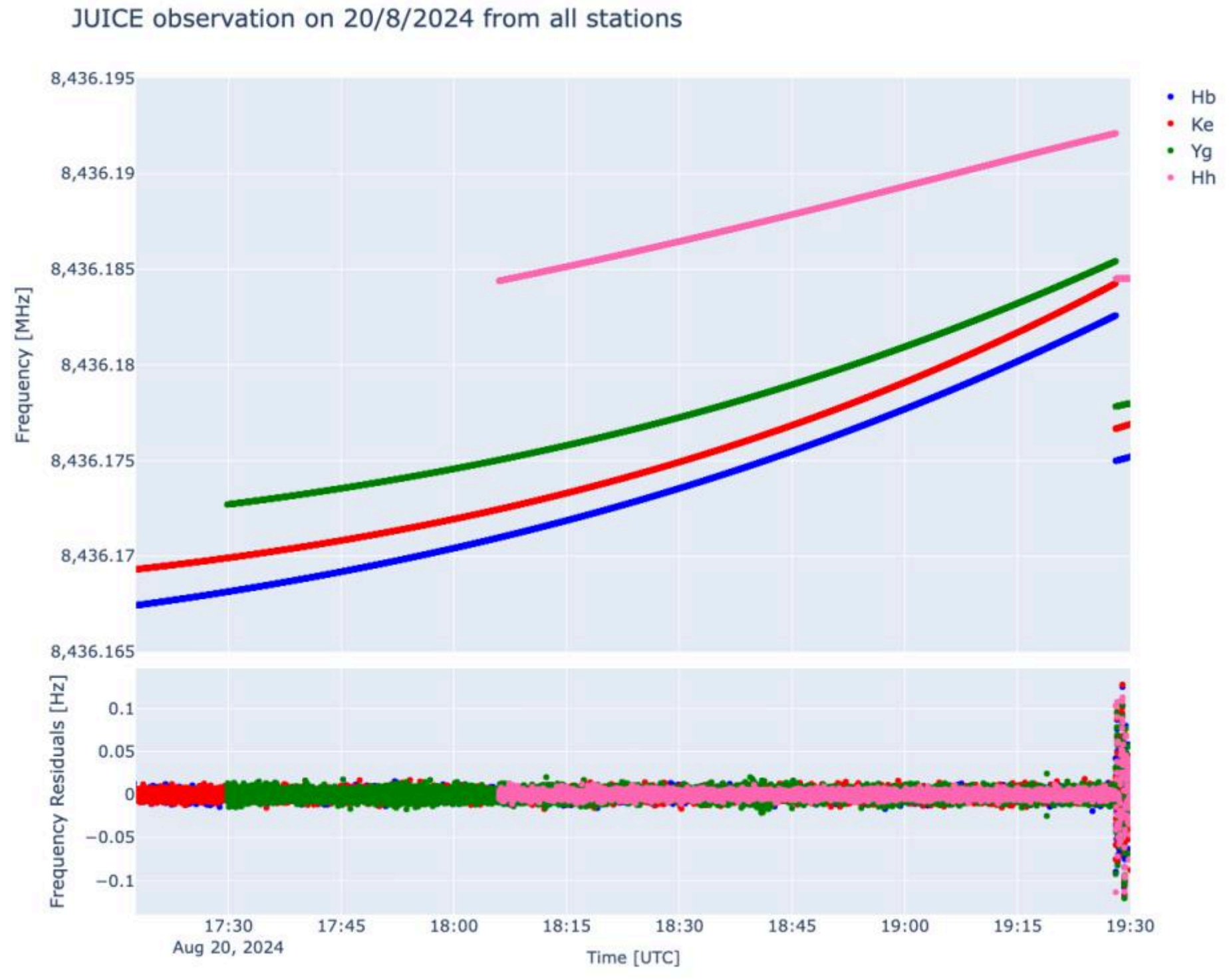

*Figure 7: Frequency detections of the JUICE flyby of the Earth as observed by Hb, Ke, Yg, and Hh. The frequency jump and increased noise in the residuals near 19:28 is due to the spacecraft switching to one-way mode.*

Analysis of the Lunar-Earth gravity assist Doppler data is ongoing with TUDAT. This is both to refine the orbital modelling of the spacecraft and test the accuracy of our Doppler data.

## Venus Flyby

On August 31st, 2025, JUICE performed a flyby of Venus. We performed PRIDE VLBI phase-referencing experiments prior to and following the flyby itself; however, due to technical issues the observations were unsuccessful. Doppler observations were conducted in the days following the flyby, as shown in Figure 8. As with the Lunar-Earth gravity assist, TUDAT will be used to analyse the data.

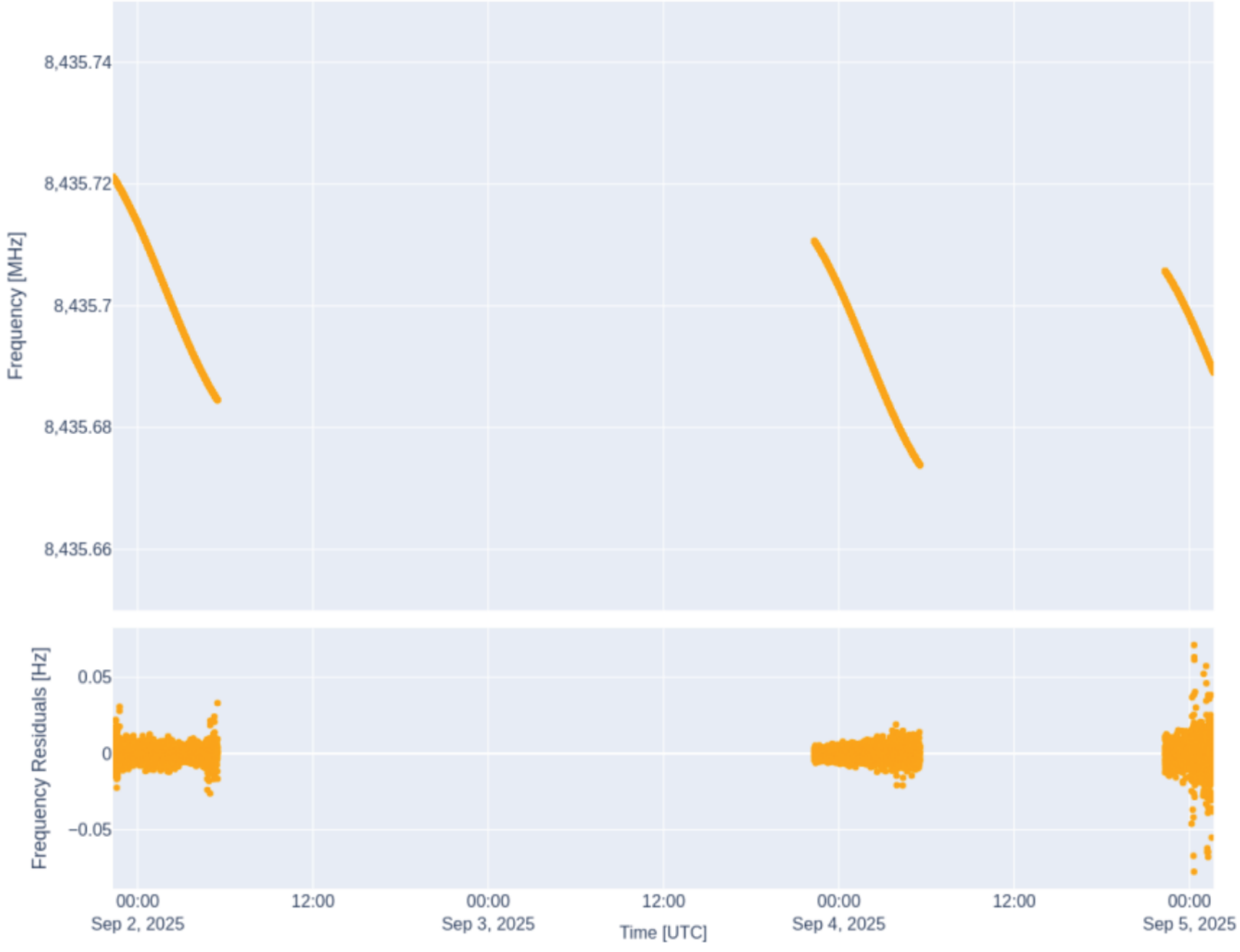

*Figure 8: Frequency detections of JUICE at Cd in the days following the Venus flyby, which transpired on August 31st, 2025.*

**Space Weather**

Of the over 100 PRIDE observations of JUICE conducted from UTAS, the vast majority have been single-dish Doppler observations, with the aim of performing scintillation analysis to determine properties of the space weather on the spacecraft signal. Table 2 shows the parameters for the observations where there were one or more scans with phases suitable for scintillation analysis.

*Table 2: Observation parameters of the PRIDE space weather observations of JUICE conducted at the University of Tasmania.*

| | Stations | Distance to Target (AU) | Solar Elongation |
|---|---|---|---|
| **20/4/2023** | Hb, Ke, Ho | 0.0097 | 156 |
| **22/4/2023** | Hb, Ke | 0.0127 | 154 |
| **24/4/2023** | Hb, Ke, Ho, Yg | 0.0157 | 152 |
| **26/4/2023** | Hb | 0.0189 | 151 |
| **2/5/2023** | Ke | 0.0282 | 146 |
| **4/5/2023** | Hb, Ke | 0.0312 | 144 |
| **8/5/2023** | Hb, Ke, Yg | 0.0376 | 141 |
| **10/5/2023** | Hb, Ke | 0.0408 | 139 |
| **11/5/2023** | Hb, Ke | 0.0426 | 139 |
| **13/5/2023** | Hb, Ke | 0.0457 | 137 |
| **14/5/2023** | Hb, Ke | 0.0476 | 136 |
| **20/5/2023** | Hb, Ke | 0.0576 | 132 |
| **21/5/2023** | Cd, Hb | 0.0593 | 132 |
| **27/5/2023** | Hb, Ke | 0.0697 | 128 |
| **2/6/2023** | Hb, Ke | 0.0804 | 124 |
| **3/6/2023** | Hb, Ke | 0.0823 | 123 |
| **8/6/2023** | Ke | 0.0912 | 120 |
| **14/6/2023** | Hb, Ke | 0.1019 | 117 |
| **19/6/2023** | Hb | 0.1110 | 114 |
| **21/6/2023** | Hb | 0.1145 | 113 |
| **22/6/2023** | Hb, Ke | 0.1163 | 113 |
| **26/6/2023** | Ke | 0.1234 | 111 |
| **5/7/2023** | Hb | 0.1402 | 106 |
| **12/7/2023** | Hb | 0.1497 | 104 |
| **14/9/2023** | Hb | 0.1932 | 77 |
| **12/10/2023** | Cd | 0.1657 | 61 |
| **12/11/2023** | Ho | 0.1221 | 25 |
| **14/8/2024** | | 0.0118 | 166 |
| **13/3/2025** | Yg | 1.3105 | 42 |
| **14/5/2025** | Cd | 1.3936 | 46 |
| **17/7/2025** | Cd | 1.2958 | 44 |
| **6/8/2025** | Cd | 1.2950 | 40 |

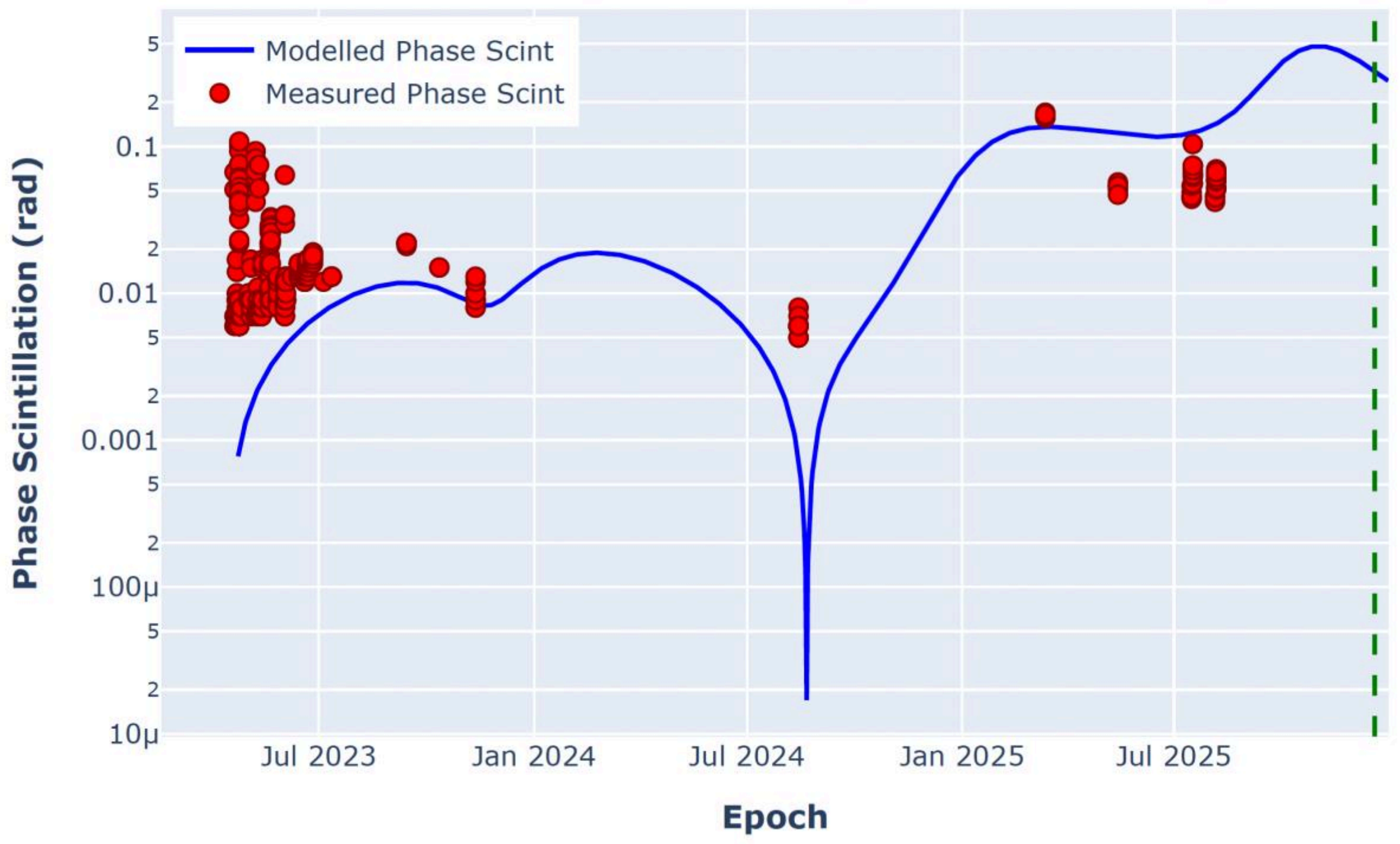

*Figure 9: Predicted versus measured phase scintillation for PRIDE single-dish observations of JUICE throughout the cruise phase. Note that some of the discrepancies are due to no corrections being applied for the ionospheric or tropospheric TEC, which will be explored in future analysis.*

We compared the measured interplanetary phase scintillation with a model based on the predicted total electron content (TEC). The nominal electron density at radial distance *r* from the Sun, in AU, is given by,

$$n_e = 5 * r^{-2} \; [cm^{-3}]. \tag{1}$$

This is assuming a nominal electron density of five electrons per cubic centimetre at 1 AU, first determined by the Mariner II probe [15]. Fast and slow solar wind models have also been fitted from data obtained in more recent missions and include higher order radial terms [16]. These models are also used in the analysis of PRIDE space weather data; however, were beyond the scope of this paper, and will be explored further in future analysis of this JUICE data.

To predict the TEC, we took points along the line of sight between the spacecraft and the Earth, with the positions predicted using coordinates from JPL Horizons[2]. At each point the nominal electron density was multiplied by the distance increment and then summed along a straight-line path from JUICE to Earth, yielding the predicted TEC along the line-of-sight. This was then converted to the predicted phase scintillation along the line-of-sight using Equation 2,

$$\sigma = TEC \times \frac{8.4\left(\frac{1200}{300}\right)^{5/6}}{2Kf_r}, \tag{2}$$

where $f_r$ is the approximate received frequency and $K$ is an empirically determined scaling constant. In previous studies of Venus Express and Mars Express a value of 2000 and 2390 respectively was used [4, 11]; in the case of Mars Express by performing least squares fitting of the long-term scintillation values. For this analysis of JUICE, 2390 was used for the model, as shown in Figure 9. These measured values are reasonably consistent with the predictions; however, the ionosphere also has a comparable contribution to the scintillation, especially at the shorter Earth-JUICE distances during the near-Earth commissioning phase.

Further analysis will involve modelling tropospheric effects, as well as using the vertical total electron content (vTEC) maps, produced by the international GNSS service, to remove the contribution of the ionosphere [10], to provide a more accurate measurement of space weather activity. Additionally, with sufficient data, an attempt at fitting a scaling constant specifically to the JUICE data will be performed, and outliers identified and analysed as potential coronal mass ejection (CME) events [13]. Finally, we will also attempt fitting electron density models that contain additional parameter dependences, such as the ecliptic latitude.

## Conclusion

The University of Tasmania VLBI-equipped radio telescopes have enabled two years of PRIDE tracking of the JUICE mission, with over 100 observations conducted. This included testing of observation modes and configurations during the near-Earth commissioning phase, with comparisons of one-way and two-way modes, and recording in different polarisations. Nine VLBI phase-referencing experiments were conducted, correlation performed successfully, and images of the spacecraft were produced. Measurements of the interplanetary phase scintillation along the line of sight between JUICE and the Earth were consistent with TEC models utilised in the analysis of similar experiments for the Mars Express and Venus Express missions.

Further PRIDE VLBI phase-referencing experiments are planned, along with refinement of the data processing pipeline to determine the precise position of JUICE in right ascension and declination. Additional phase scintillation data will be collected throughout the remainder of the cruise phase, and ionospheric effects will be removed to improve the comparison with theoretical models. These models will also be expanded to further reflect the complexity of the solar system plasma environment.

[2] https://ssd.jpl.nasa.gov/horizons/

## Acknowledgements

This work was supported by an Australian Government Research Training Program (RTP) Scholarship.